\documentstyle[epsfig]{mn}
\def\cm{\mbox{\,cm}}

\def\s{\mbox{\,s}}

\def\tento#1{\times 10^{#1}}

\def\K{{\rm \ K}}

\def\gtsima{$\; \buildrel > \over \sim \;$}
\def\ltsima{$\; \buildrel < \over \sim \;$}
\def\prosima{$\; \buildrel \propto \over \sim \;$}
\def\gsim{\lower.7ex\hbox{\gtsima}}
\def\lsim{\lower.7ex\hbox{\ltsima}}
\def\simgt{\lower.7ex\hbox{\gtsima}}
\def\simlt{\lower.7ex\hbox{\ltsima}}
\def\simpr{\lower.7ex\hbox{\prosima}}

\def\edcomment#1{\iffalse\marginpar{\raggedright\sl#1\/}\else\relax\fi}
\reversemarginpar



\title[Radiative Transfer around Point
 Sources]{Adaptive Ray Tracing for Radiative Transfer around Point
 Sources}
 \author[T. Abel and B. D. Wandelt]{Tom Abel$^{1,2}$
 and Benjamin D. Wandelt$^{3,4,5}$ \\
$^1$ Harvard Smithsonian Center for Astrophysics, Cambridge, MA, 02138,
USA
 \\
$^2$ Institute of Astronomy, Madingley Road, Cambridge, CB03 0HA, UK
\\
$^3$ Department of Physics, Princeton University, Princeton, NJ 08544,
USA\\
$^4$ Department of Physics, University of Illinois, Urbana, IL 61801,
USA\\
$^5$ Department of Astronomy, University of Illinois, Urbana, IL 61801,
USA }
\date{revised manuscript of a Letter to the Editor for publication in
MNRAS}
\pubyear{2001}
\begin{document}
\label{firstpage}
\maketitle

\begin{abstract}
  We describe a novel adaptive ray tracing scheme to solve the equation of
  radiative transfer around point sources in hydrodynamical simulations. The
  angular resolution adapts to the local hydrodynamical resolution and hence
  is of use for adaptive meshes as well as adaptive smooth particle
  hydrodynamical simulations.  Recursive creation of rays ensures ease of
  implementation. The multiple radial integrations needed to solve the time
  dependent radiative transfer are sped up significantly using a quad--tree
  once the rays are cast.  Simplifications advantageous for methods with
  one radiation source are briefly discussed. The suggested method is easily
  generalized to speed up Monte Carlo radiative transfer techniques.  In
  summary a nearly optimal use of long characteristics is presented and
  aspects of its implementation and comparison to other methods are given.
\end{abstract}

\section{Introduction}

Numerous questions in physical cosmology and galaxy formation require
a detailed understanding of the physics of radiative transfer.  Of
particular interest are problems related to the reionization of the
intergalactic medium (Haiman, Abel, and Madau 2000, Barkana and Loeb
2000, and references therein), absorption line signatures of high
redshift structures (Kepner et al. 1999; Abel and Mo 1998).
Hydrodynamic and thermal effects of UV emitting central sources on
galaxy formation are also of interest (e.g. Haehnelt 1995; Silk and
Rees 1998).  

Consequently a number of methods for solving radiative transfer in
three dimensional cosmological hydrodynamical simulations have been
developed. Abel, Norman and Madau (1999, ANM99 hereafter) introduced a
ray tracing scheme that employed rays of uniform angular resolution to
solve the time dependent radiative transfer on uniform Cartesian grids
for point sources.  A related ray tracing approach was introduced by
Razoumov and Scott (1999) which can also handle diffuse radiation
fields but limited computer memory limits the angular resolution and
therefore only poorly captures ionization fronts around point sources
on reasonably large grids. Ciardi et al. (2001) discuss a Monte Carlo
approach to sample angles, energies and the time evolution of the
radiative transfer problem. Such Monte Carlo techniques may be able to
use aspects of our new method to choose a more optimal number of
photon packets. A ray--tracing scheme for use with Smooth Particle
Hydrodynamics was introduced by Kessel-Deynet and Burkert (2000).
Moment methods using variable Eddington tensors in three dimensions
have been suggested by (Norman, Paschos, and Abel 1998) and
implemented (Gnedin and Abel 2001). Please note that there is much
more literature on the problem of radiative transfer in two dimensions
(c.f. Turner and Stone 2001, and references therein).

In this Letter we introduce a novel adaptive ray tracing scheme which
is applicable for solving any line integrals on radial rays on uniform
and adaptive Cartesian grids as well as particle methods.

The algorithm and aspects of an implementation are explained in the
following section. Section~\ref{ex} gives an illustrative example
application. In the discussion section~\ref{disc} we outline
strategies to adapt our method to some particular applications of
interest.

\section{Method}
\subsection{Illustrative Problem}
To motivate and introduce the new method we consider the specific
example problem of solving the monochromatic radiative transfer of
Lyman continuum photons from an isotropic point source photo--ionizing
a neutral hydrogen density distribution, $n_H(\vec{r},t)$, given on a
uniform Cartesian grid. The equation of radiative transfer for the
photon number flux $N_{P}$ [\#/s] along a ray reads
\begin{eqnarray}
\frac{\partial N_{P}}{c \partial t} + \frac{\partial N_P }{\partial r} =
 - \kappa N_P,
\end{eqnarray}
where the absorption coefficient $\kappa = \sigma n_H(\vec{r},t)$ is given
from cross--section for photoionization $\sigma \sim 6.3\tento{-18}\cm^{-2}$
at threshold. For simplicity we consider here only photons at
threshold. Including the frequency dependency is easily achieved repeating the
ray tracing. If the light crossing times are shorter than the opacity time
scale $\kappa/\dot{\kappa}$ the time dependent term can be dropped and only
the static attenuation equation needs to be solved (ANM99; Norman, Paschos,
Abel 1998). The rate of photo--ionizations, $k_P^i(\vec{r})$ [s$^{-1}$
cm$^{-3}$], contributed by the {\em i}th ray carrying the photon number flux,
$N_P^i$, and passing through a cell with neutral hydrogen density,
$n_H(\vec{r})$, for a length, $\triangle r^i$, is given by $k_P^i(\vec{r}) =
N_P^i (1-e^{-\tau^i})/(\Delta x^3) = N_P^i [1-\exp(- n_H(\vec{r}) \sigma \triangle r^i)]/(\Delta x^3)$.
The total rate of photo--ionization for a given cell is sum of all rays
passing through that cell. The number of rays passing through a cell controls
the accuracy of the angular integration. Any ray tracing scheme requires of
the order $ 4\pi R^2/(\triangle x) ^2$ rays to have at least one ray pass a
cell of side length $\triangle x$ at a distance $R$ from the source. If the
number of rays per solid angle is fixed as one traverses the simulation
volume, one integrates through cells close to the source unnecessarily more
often in order to resolve cells farther away.  A similar problem is inherent
in Monte Carlo techniques that send photon packages and have to ensure
sufficient angular resolution (Ciardi et al. 2001, and references therein).

Using an {\em adaptive} number of rays overcomes these limitations. We
implement this adaptivity by growing a  tree of rays out of the point
source, subdividing  rays locally as they encounter grid cells. The
tree can be recombinant, with  refinement or coarsening 
occurring to ensure all cells which are illuminated by the source are
traversed by a sufficient number of 
rays. In the following we discuss how ray   
directions are chosen.

\subsection{HEALPix Pixelization of the sphere}

We use the Hierarchical Equal Area isoLatitude Pixelization  of the
sphere (HEALPix). The
details of this pixelization are described in G\'orski, Hivon, and
Wandelt (1998). 
For our purposes, its advantageous properties are, a) exactly equal area
per pixel, b)
nearly uniform sampling of solid angle by the pixel centers, c)
hierarchical nesting of pixels at different resolutions, and d) the
binary {\bf nested} pixel numbering scheme with fast routines provided
for converting between pixel numbers and the associated unit vectors.

Figure 1 shows how HEALPix provides a
tiling of the sphere with these properties at arbitrarily high
resolution by recursively subdividing a set of
12 base pixels. At each refinement step the pixels divide
into 4  next-level pixels and thus generate a
quad-tree structure. At the  
$l$-th level ($l=0,1,...$), the tiling contains $N_{pix}(l)=12\times 4^l$
pixels each of which covers solid angle $A(l)=4\pi/N_{pix}(l)$. This
equal area property ensures equal photon flux per ray for
isotropic sources. 

This HEALPix quad--tree allows a recursive implementation of an
adaptive algorithm for defining a ray tree.
Individual rays split into $4^n$ ($n=1,2,...$) child
rays when necessary to preserve the accuracy of the angular integration. 
Adopting the {\bf nested} numbering scheme, the
numbers of child rays 
are simply given by adding two bits to the number of the parent rays.
This has advantages in memory management and implementation
as we will discuss in more detail below. 

\vspace{.3cm}
\begin{figure}
\vspace{-1cm}
\centerline{\psfig{file=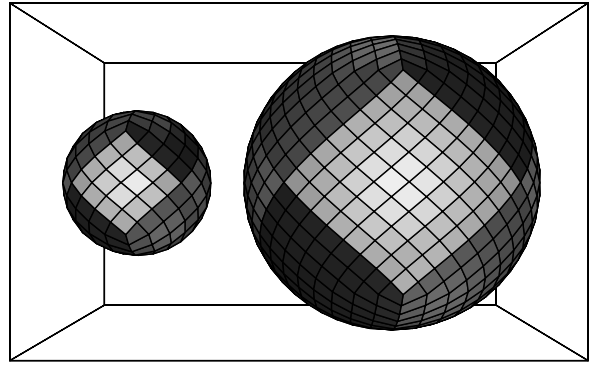,width=.45\textwidth}}
%\vspace{-1cm}
\caption{\footnotesize{HEALPix pixelization of the sphere. The 12 base
pixels are coloured in and shown at resolution $l$=2 (left) and $l$=3 (right)
Rays are
cast through pixel centers. The hierarchical property of the
pixelization enables the
efficient construction of a tree of rays (see Fig.~2). }}\vspace{-.2cm}
\end{figure}

\subsection{Splitting Rays and Setting up the Tree}
As one walks along a ray a refinement criterion controls whether this ray
needs to be broken up into 4 child rays. For the uniform grid the refinement
is controlled by keeping $A_c/A(l)>f$. This is the ratio of the area of a cell
$A_c=\triangle x^2$ and the area associated with the ray on level $l$,
$A(l)=4\pi/(12 \times 4^l)$. Here we introduced a safety factor $f\gg 1$ which
is the minimum number of rays passing through any cell. We find $f=2$ to be a
good compromise between speed and accuracy.  For opacities given on a
structured adaptive mesh this refinement criterion is unchanged provided one
uses the correct local cell area. For an adaptive smooth particle
hydrodynamical representation of the opacities one would use a fraction of the
square of the local smoothing length as refinement criterion.  The path length
through the particle is the line integral through the smoothing kernel.

Once one has decided to split the current ray one determines the four new
pixel (ray) numbers by adding two bits to the current ray number. Given the
pixel number the HEALPix {\tt pix2vect\_nest} routine is used to find the new
unit directional vectors for the child rays. The child ray starting points are
simply given by the radius of the parent ray times the directional vector. One
subtlety is that these starting points may reside in a different cell of the
Cartesian grid and hence the associated cell number needs to be recomputed. A
tree is realized using two pointers for each ray. We refer to them as {\tt
  NextRayThisLevel} and {\tt NextRayNextLevel}. As a parent ray splits into
child rays it will have {\tt NextRayNextLevel} pointing at the ray number of
the first child ray. The first three child rays are created with setting {\tt
  NextRayThisLevel} to the ray numbers according to their pixel numbering. The
last child ray is created with {\tt NextRayThisLevel} pointing to a negative
integer value indicating the end of the tree on that level. Given these two
pointers per each ray one can quickly walk through the tree of rays.  When the
ray is split the photon--number flux is equally distributed to the child rays.

\begin{figure}
\epsfysize=10cm
\epsfxsize=12cm
\centerline{\psfig{file=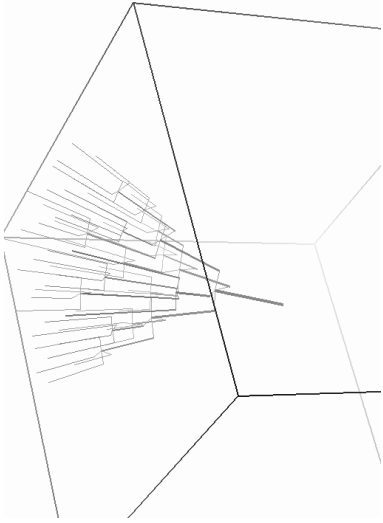,width=5cm}}
\caption{Illustration how one base ray spawns child rays
to sample a uniform underlying grid preserving the accuracy of the
angular integrations.  }\label{1ray}
\end{figure}

Figure~\ref{1ray} gives a graphical illustration of how one base ray
is split up into child rays on three additional levels in a uniform
grid with only 8 cells on a side.     
\vspace{.2cm}

\subsection{Merging of Rays}

For ray-tracing through opacity data with a large dynamic spatial range
situations may arise in which it is useful to merge child rays again and start
following a new ray of the direction of their parent ray. For a non-uniform
Cartesian grid such a criterion is readily specified by again comparing the
area associated with the ray to the area associated with the local opacity
grid.  By keeping $A_c/A(l)<f_M$ one never traces more than $f_M$ rays through
any cell. To set up the tree with the merging algorithm one now can set {\tt
  NextRayNextLevel} of the last of the four child (now parent) rays to point
to the merged ray number. {\tt NextRayThisLevel} of that last child (now
parent) ray remains set to a negative integer value. If merging is used one
also needs to define a flag that remembers that it is merged ray. This flag is
needed to evaluate the incoming photon flux of the merged ray from the sum of
the outgoing photon fluxes of the 4 parent rays.  This setup strategy of the
tree is simple to implement, but conceptually not very elegant since merged
rays are labeled at a higher level than they really are. However, since this
does not affect the efficiency of the integration we do not discuss this here
further.

\subsection{Ray Ending criteria}
In many problems of radiative transfer one can define a simple criterion that
justifies to stop tracing rays any further. In our example, the propagation of
ionization fronts, it is obvious that one does not need to trace rays after
most (e.g. 99\%) of the photon flux ($N_P/(12 \times 4^l$) associated
with a ray has been absorbed.  This leads to a dramatic reduction of
computing expense if only a small fraction of the volume is affected
by the radiation. In general, which ray ending criterion is suitable
depends on the application at hand.

\subsection{Integrating along Rays}
The radial integration on Cartesian grids is done identical as explained in
detail in ANM99 and we discuss it here only briefly. One simply finds the
intersections with the 6 planes containing the faces of the next cell along
the x, y, and z coordinates and computes the length of ray segments. The
shortest ray segment is the relevant one. In our example problem one then
knows the optical depth and hence the photon number flux emerging after the
ray has transversed the current ray segment. This in turn gives the
corresponding photo ionization rate which is then added to the total
photo-ionization rate on the Cartesian grid.  In situations where the same
radial integration has to be performed multiple times we find it advantageous
to store the length of the ray segments and the associated opacity grid
indices for each ray.

\subsection{Walking through the Tree}
One major advantage of the proposed method is the inherent tree
structure which allows quick multiple integrations once the tree has
been defined.  Given the pointers {\tt NextRayThisLevel} and {\tt
NextRayNextLevel} which were introduced when the rays were defined the
walking through the tree is simple. It is programmed by recursively
calling a function that will first integrate the ray then call itself
to integrate the ray given by {\tt NextRayNextLevel} pointer and after
that the ray recorded by {\tt NextRayThisLevel}. Before, one calls
this function to integrate {\tt NextRayNextLevel} one copies the
integral quantities (e.g. the photon flux of the ray) to all its
child--rays. Alternatively the routine before integrating a ray could
copy the integral quantity from its parent. The ray data structure
will also need to store a pointer to the parent ray.

\section{Example}\label{ex}

We have tested our scheme on all cases presented in ANM99 and found
the results indistinguishable from their non adaptive ray tracing. As
a new illustrative example we integrate the radial jump condition of
an ionization front in three dimensions. 

Consider the simple case of a constant luminosity point source
embedded in a static medium of neutral hydrogen. The jump condition
along a ray at the location of the ionization front caused by the
source reads
\begin{eqnarray}
\frac{N_P}{4\pi R^2} - \int_0^{R} n^2 \alpha(T) dr = n \frac{dR}{dt}.
\end{eqnarray}
Here, $n$, denotes the number density of hydrogen nuclei, $\alpha(T)$,
the (in general) temperature dependent recombination rate coefficient,
$R$, the location of the ionization front and $N_P$ the ionizing
photon number flux of the source. If we assume the ionized material to
have a constant temperature and define $\alpha_4=\alpha(10^4\K)$ one
can integrate this equation with first order differencing to find the
passing time, $t_p(R)$. The ray tracing gives one a list of $\triangle
R_i$ ray segments from which one finds $t_p$ via
\begin{eqnarray}
t_p^{i+1} = t_p^i + \triangle R_i\,\, n(R) \left[\frac{N_P}{4\pi R^2} -
\int_0^{R} n^2\, \alpha_4\, dr \right]^{-1}. 
\end{eqnarray}
The integral of the recombinations from the source to the ionization front is
evaluated on the fly. Via this technique one finds the entire evolution of the
ionization front in one radial integration.  The passing time through a cell
is taken to be the maximum passing time evaluated from all rays passing the
cell. The classical jump condition gives in the limit of small $R$ a speed of
the ionization front $dR/dt$ exceeding the speed of light. To avoid
unphysically fast expansion times we therefore limit the maximum $\triangle
t_p$ to be the light crossing time $\triangle R/c$, where, $c$, denotes the
speed of light. The computation is carried out on a uniform Cartesian grid
containing 128$^3$ cells with the source at one of the corners of the grid. We
start with $l=2$ giving 192 base rays. To evaluate all arrival times
$1.4\tento{5}$ rays were used. We used a homogeneous initial neutral hydrogen
density of $1\cm^{-3}$ with a cubic obstacle of 10 times larger density and a
source with a photon--luminosity of $N_P=10^{47}\s^{-1}$.
\begin{figure}
\epsfysize=10cm
\epsfxsize=12cm
\centerline{\psfig{file=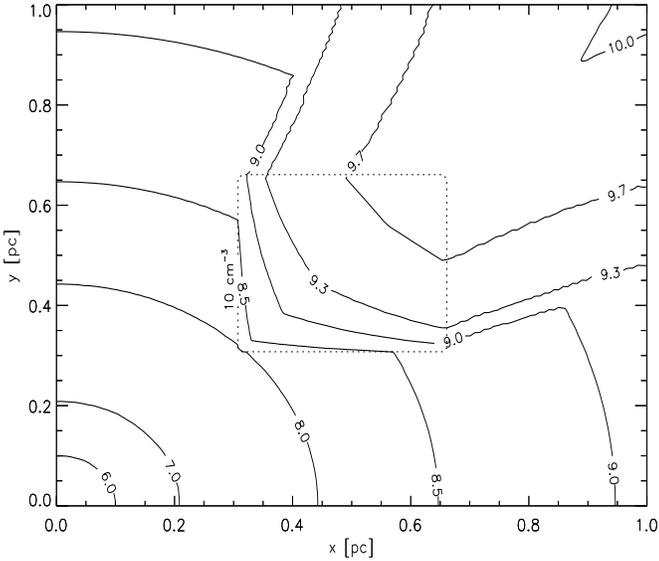,width=9cm, height=8cm}}
\caption{\footnotesize{Source of $N_P=10^{47}\s^{-1}$ embedded in a
    uniform medium of  $n=1\cm^{-3}$ and a cubic obstacle  of ten times higher
    density ($n=10\cm^{-3}$). The contours give the location of the ionization
    front   at  $10^6$,  $10^7$,   $10^8$,  $10^{8.5}$,   $10^9$,  $10^{9.3}$,
    $10^{9.7}$ and $10^{10}$ seconds after  the source switched on. The dotted
    line  marks   the  location  of  the  obstacle.   The  simulations  volume
    corresponds to side length of one parsec.}}\label{con}
\end{figure}
Contours  in  Figure~\ref{con}  indicate  the  ionization  front  location  at
different  times.  A  two dimensional  slice of  the three  dimensional volume
through the position of the source is shown. The perfect shadowing is evident.
Compared to non--adaptive ray tracing this new method is about 20 times faster
in  this  test  problem.  Using  the  tree for  the  integration  led  in  our
implementation to a further speed--up of  a factor of two. Note, however, that
larger speed--ups  will be  achieved in situations  with large  dynamic range,
where one can stop rays that enter optically thick regions.

\section{Discussion}\label{disc}

In  summary we  have presented  a novel  approach to  simulate the  effects of
radiative transfer  around multiple  point sources. The  use of  adaptive rays
with a an inherent tree  structure allows significant speed--ups in comparison
to uniform  ray tracing.  Although we have  developed it with  applications to
astrophysical hydrodynamics  in mind  it is equally  well suited  for studying
static  situations and  may  find  application also  for  volume rendering  in
computer graphics. The  method may also be used  for extended emitting sources
which are then modeled by a collection of point sources.

It seems clear that generalization of this method may be exploited to speed up
Monte  Carlo  radiative  transfer  techniques  (e.g. Ciardi  et  al  2001  and
references therein). Here one would  start with fewer luminous photon packages
at  the source  and split  up the  the packages  to ensure  similar  number of
packages to  transverse the numerical  grid on which  the opacity is  defined. 
This should speed up the calculations significantly since again cells close to
the source need not to be accessed more often than any other grid cells.

The adaptive ray tracing scheme suggested in this Letter, allows fast multiple
integrations in the case of non-moving sources using a quad--tree. Also, for a
variety of problems one may not always  have to trace all rays from the source
to all  other grid cells (or  particles). For example,  in the case of  a thin
ionization front  penetrating into a  high density medium the  opacity changes
only significantly at the  front. Rays need not to be traced  far ahead of the
front  where most  of their  photons have  been consumed.   Rays close  to the
source may  also not needed to be  traced if their (already  low) opacity does
not change  significantly. Based on this  one can formulate a  scheme in which
rays carry  their own opacity time  step. As one integrates  the chemical rate
equations one  traces only rays  and their child  rays which carry  an opacity
time step less or equal to the current chemical times step. Such a scheme that
is also adaptive in time is  necessarily more complex and needs to be tailored
to the specific problem at hand as well as the method employed for solving the
hydrodynamics. We  are currently  exploring such time  adaptive schemes  to be
implemented  for  smoothed  particle  as  well  as  structured  adaptive  mesh
refinement techniques.

\section*{Acknowledgments}
T.A. acknowledges  stimulating and insightful  discussions with Greg  Bryan on
adaptive  numerical   schemes  and  Benedetta  Ciardi  for   comments  on  the
manuscript.  This  work  has  partially  been  supported  through  NSF  grants
ACI96-19019  and AST-9803137.  T.A. also  acknowledge support  from  the Grand
Challenge  Cosmology Consortium.  B.D.W. is  supported by  the  NASA MAP/MIDEX
program.

\end{document}